\documentclass[aps,prl,twocolumn,groupedaddress,amsmath,amssymb,nofootinbib,showpacs]{revtex4-1}
\usepackage{graphicx}  % needed for figures
\usepackage{dcolumn}   % needed for some tables
\usepackage{bm}        % for math
\usepackage{color}
\allowdisplaybreaks
\begin{document}
\def\be{\begin{equation}}
\def\ee{\end{equation}}
\def\bea{\begin{eqnarray}}
\def\eea{\end{eqnarray}}
\def\ba{\begin{array}}
\def\ea{\end{array}}
\def\ben{\begin{enumerate}}
\def\een{\end{enumerate}}
\def\nab{\bigtriangledown}
\def\tpi{\tilde\Phi}
\def\nnu{\nonumber}
\newcommand{\eqn}[1]{(\ref{#1})}
\def\bw{\begin{widetext}}
\def\ew{\end{widetext}}
\newcommand{\half}{{\frac{1}{2}}}
\newcommand{\vs}[1]{\vspace{#1 mm}}
\newcommand{\dsl}{\pa \kern-0.5em /} 
\def\a{\alpha}
\def\b{\beta}
\def\g{\gamma}\def\G{\Gamma}
\def\d{\delta}\def\D{\Delta}
\def\ep{\epsilon}
\def\et{\eta}
\def\z{\zeta}
\def\t{\theta}\def\T{\Theta}
\def\l{\lambda}\def\L{\Lambda}
\def\m{\mu}
\def\f{\phi}\def\F{\Phi}
\def\n{\nu}
\def\p{\psi}\def\P{\Psi}
\def\r{\rho}
\def\s{\sigma}\def\S{\Sigma}
\def\ta{\tau}
\def\x{\chi}
\def\o{\omega}\def\O{\Omega}
\def\k{\kappa}
\def\pa {\partial}
\def\ov{\over}
\def\nn{\nonumber\\}
\def\ud{\underline}
\def\ct{\textcolor{red}{\it cite }}

%\nofiles

%\preprint{APS/123-QED}

\title{\large{\bf Decoupling limit and throat geometry of non-susy D3 brane}}
\author{Kuntal Nayek}
\email{kuntal.nayek@saha.ac.in}
\affiliation{
   Saha Institute of Nuclear Physics,\\
   1/AF Bidhannagar, Kolkata 700064, India
   }
\author{Shibaji Roy}
\email{shibaji.roy@saha.ac.in}
\affiliation{
   Saha Institute of Nuclear Physics,\\
   1/AF Bidhannagar, Kolkata 700064, India
   }
\date{\today}

\begin{abstract}
Recently it has been shown by us that, like BPS D$p$ branes,
bulk gravity gets decoupled from the brane even for the non-susy D$p$ branes 
of type II string theories indicating a possible extension of AdS/CFT correspondence
for the non-supersymmetric case. In that work, the decoupling of gravity on the non-susy D$p$ branes
has been shown numerically for the general case as well as analytically for some 
special case. Here we discuss the decoupling limit and the throat geometry of the 
non-susy D3 brane when the charge associated with the brane is very large. We show that
in the decoupling limit the throat geometry of the non-susy D3 brane, under appropriate coordinate
change, reduces to the
Constable-Myers solution and thus confirming that this solution is indeed the holographic
dual of a (non-gravitational) gauge theory discussed there. We also show that when
one of the parameters of the solution takes a specific value, it
reduces, under another coordinate change, to the five-dimensional solution obtained by Csaki and Reece, again confirming
its gauge theory interpretation.\end{abstract}

%\pacs{11.25.-w, 11.25.Tq, 11.25.Uv}

% PACS, the Physics and Astronomy
                             % Classification Scheme.
%Use showkeys class option if keyword
                              %display desired
\maketitle

AdS/CFT correspondence, in its original version \cite{Maldacena:1997re} (see also \cite{Witten:1998qj,Gubser:1998bc}), is an 
equivalence or a duality between two theories --
one is a non-gravitational, conformally invariant and supersymmetric field theory in four dimensions
(more precisely, $D=4$, ${\cal N}=4$ super Yang-Mills theory) and the other is a string theory (type IIB)
or a gravitational theory in AdS space in five dimensions (times a five-dimensional sphere). It is 
holographic and is a strong-weak duality symmetry, in the sense, that when the field theory is strongly
coupled the string theory is weakly coupled (given by supergravity) and vice-versa. This duality, is
therefore, very useful to understand the strong coupling behavior of field theory by studying the weakly
coupled string theory or supergravity. However, the theories on both sides of this duality are supersymmetric
as well as conformally invariant and therefore not very realistic like QCD theory which
is neither supersymmetric nor conformal. AdS/CFT correspondence has been extended for the less 
supersymmetric \cite{Klebanov:1998hh}, non-conformal cases \cite{Klebanov:2000nc,Klebanov:2000hb} and even 
in other dimensions (other than three) \cite{Itzhaki:1998dd} generally known as 
gauge/gravity duality (see \cite{Aharony:1999ti} for a review). AdS/CFT type correspondence has also been studied for the 
non-supersymmetric (type 0) string theory solutions in \cite{Klebanov:1998yya}. 

There is no doubt that AdS/CFT type correspondence will be more useful if it can be understood for the
non-supersymmetric, non-conformal case, where the associated field theory would be more like QCD and various strong
coupling behavior of QCD can be understood by studying the dual gravity theory. However, the exact dual
gravity theory which would correspond to QCD on the boundary is not known. But, it is clear that
the relevant gravity solution must be non-supersymmetric. So, one could either start with a BPS 
brane-like solution and break the supersymmetry by compactification \cite{Witten:1998zw} or start directly with the 
non-supersymmetric brane-like solution of type II string theory \cite{Zhou:1999nm}. Now, for gauge/gravity
duality to work in a brane-like gravitational background, there must exist a low energy or a decoupling limit
for which bulk graviton must decouple from the brane. This can be shown either by calculating the graviton potential
in the brane background which takes the form of an infinite barrier or by calculating the graviton absorption 
cross-section which vanishes in the decoupling limit.
This is precisely what happens for the BPS D$p$ branes of type II string theory
\cite{Das:1996wn,Klebanov:1997kc,Gubser:1998iu,Alishahiha:2000qf} and in \cite{Nayek:2015tta}, we
have shown that exactly the same phenomenon occurs for the non-supersymmetric D$p$ brane solutions
of the same theory as well. Usually it is assumed that gauge/gravity duality should work even for
the non-supersymmetric case and the results in \cite{Nayek:2015tta} clearly indicate that this is indeed true. 

In this Letter we will consider the non-supersymmetric D3 brane solution of 
type IIB string theory and work out the decoupling limit more clearly. (An anisotropic non-susy D3 brane solution 
has been shown, by zooming into a particular space-time
region, similar to the decoupling limit discussed in this Letter, to interpolate between AdS$_5$ black hole, AdS$_5$
soliton and a soft-wall gravity solution in \cite{Roy:2015zka}.) 
In obtaining the decoupling limit for the 
non-susy case we will draw analogy from the BPS case and make sure that the decoupling limit goes over to the 
BPS D3 brane decoupling limit, when susy is restored. We will also show that the low energy excitations in the 
throat region and in the bulk get decoupled in the decoupling limit from
the energy considerations. We then give the throat geometry which keeps the effective string action
finite. Finally, by making an appropriate coordinate transformation, we show that the geometry is
actually identical with the two parameter solution obtained previously by Constable and Myers \cite{Constable:1999ch}. 
We further show that when we fix one of the parameters and make
another coordinate transformation the geometry reduces precisely to the one studied by Csaki and 
Reece \cite{Csaki:2006ji}. Thus our result justifies the gauge theory interpretation
due to decoupling of gravity on the brane.  

{\em Non-susy D$3$ brane} -- The form of the non-supersymmetric D$3$ brane solution of type IIB string theory 
can be obtained by putting $p=3$ in eq.(2.11) of \cite{Nayek:2015tta} which is
\bea\label{nonsusyd3n}
ds^2 & = & F(\rho)^{-{\frac{1}{2}}}G(\rho)^{\frac{\delta}{4}}\left(-dt^2+\sum_{i=1}^3 (dx^i)^2\right)\nn
&&+F(\rho)^{\frac{1}{2}}G(\rho)^{\frac{1+\delta}{4}}\left(\frac{d\rho^2}{G(\rho)}+\rho^2 d\Omega^2_5\right),\nn
e^{2\phi} & = & G(\rho)^\delta,\quad F_{[5]} = \frac{1}{\sqrt{2}}(1 + \ast) Q {\rm Vol}(\Omega_5)
\eea
where, the functions $F(\rho)$ and $G(\rho)$ are given as,
\bea\label{functions}
F(\rho) & = &  G(\rho)^{\frac{\alpha}{2}} \cosh^2\theta - G(\rho)^{-\frac{\beta}{2}} \sinh^2\theta,\nn
G(\rho) & = & 1+\frac{\rho_0^4}{\rho^4},
\eea 
In the above the metric is given in the string frame and we have suppressed the string coupling
constant $g_s$ which is assumed to be small. The metric in \eqref{nonsusyd3n} has SO(1,3) $\times$ SO(6) symmetry
and so, the solution is not of ``black brane'' type, rather, it is of BPS type. A ``black brane'' type solution should have
R $\times$ SO(3) $\times$ SO(6) symmetry. $F_{[5]}$ is the self-dual RR 5-form and $Q$ is the charge of the non-susy D3 brane.
Note from \eqref{functions} that because of the form of $G(\rho)$, the solution has a naked singularity at $\rho=0$ and
the physical region is given by $\rho>0$. Further note that the solution is characterized by six parameters, namely, $\a$, $\b$,
$\d$, $\theta$, $Q$ and $\rho_0$ of which $\rho_0$ has the dimension of length, $Q$ has the dimension of four-volume 
and others are dimensionless. Since $e^{\phi}$ is the effective string coupling, the gravity solution \eqref{nonsusyd3n} will remain valid
only when the parameter $\d$ is less than or equal to zero and the radius of curvature (in string units) associated with the solution is
very large. This latter restriction is satisfied in the decoupling limit when the charge of the brane is very large as discussed
in the next section.    
The parameters of the solution mentioned above are not all independent as they
must satisfy certain constraints for the consistency of the equations of motion. The constraints are,
\bea\label{constraints}
& & \a = \b,\quad\quad Q = 2 \a \rho_0^4 \sinh2\theta,\nn
& & \a^2 + \d^2 = \frac{5}{2} \quad \Rightarrow -\sqrt{\frac{5}{2}} \leq \d \leq 0
\eea
Note that the above non-supersymmetric D$3$-brane is asymptotically flat. We can compare the non-susy D3 brane solution
\eqref{nonsusyd3n} with the BPS D3 brane solution. First of all, note that the non-susy D3 brane solution \eqref{nonsusyd3n}
contains three independent parameters ($\rho_0$, $\theta$, $\delta$), whereas BPS D3 brane contains only one parameter (even the
black D3 brane contains two parameters). Also BPS D3 brane is always charged under RR form-field, but the non-susy
D3 brane can be chargeless by either putting $\theta$ or $\alpha$ (which is related to $\delta$ by \eqref{constraints})
or both to zero (see \eqref{constraints}). Finally, we note that for non-susy D3 brane, the dilaton is in general
not constant, however, it can be made constant by setting $\delta$ to zero. But since $\alpha$ and $\delta$ are related by
\eqref{constraints}, they can not be simultaneously put to zero.      

We can recover BPS D3 brane solution from the non-susy D3 brane solution given in \eqref{nonsusyd3n} using a double scaling
limit $\rho_0 \to 0$, $\theta \to \infty$, such that $(\a/2)\rho_0^4(\cosh^2\theta + \sinh^2\theta) \to R^4$ = fixed. Under this limit
$G(\rho) \to 1$, and $F(\rho) \to (1+R^4/\rho^4)$ and $Q \to 4R^4$ and then the solution \eqref{nonsusyd3n} reduces to standard BPS D3
brane solution.

We would like to remark that as the solution given in \eqref{nonsusyd3n} is not supersymmetric and has a naked singularity at
$\rho=0$, it is quite natural to ask whether the solution is stable under small classical perturbations. Unfortunately, the 
answer to this question with its full generality is not known. The study of stability under linear perturbations of non-supersymmetric 
space-time such as the Schwarzschild black holes both in four and higher dimensions has a long history and are given in \cite{Regge:1957td}.
These studies have been extended even for the globally naked singular solution in four and higher dimensions in \cite{Gibbons:2004au,Ishibashi:2004ds} 
and for
the black $p$-brane solutions in higher dimensions in \cite{Gregory:1993vy}. Keeping in mind the cosmic censorship hypothesis one might think
that globally naked singular solution, such as the one discussed in this Letter, must be unstable under linear perturbations, but careful 
analysis given in \cite{Gibbons:2004au}, suggests that this apprehension is not always correct and there are stable nakedly singular solutions for 
certain physical boundary conditions. This has also been corroborated in the study of \cite{Sadhu:2012ur}.        

In \cite{Nayek:2015tta}, we have studied the dynamics of small classical graviton perturbations of scalar type (i.e., the perturbations are
along the brane) and obtained a Schr\" odinger type equation satisfied by it. The analysis of the
potential in this case suggests that at least for the scalar perturbations the background is stable. However, to claim that
the space-time \eqref{nonsusyd3n} is stable under linear perturbations we must also study the vector as well as the tensor
perturbations with the proper boundary condition at the singularity \cite{Ishibashi:2004ds}. This problem is currently under investigation.
    
{\em Decoupling limit} -- As we know the decoupling limit is a low energy limit by which the fundamental string length $\ell_s = \sqrt{\a'} \to 0$.
In this limit not only the interactions between the bulk theory and the theory living on the brane vanish, but also all the
higher derivative terms in both the theories go to zero. Also as we have seen \cite{Nayek:2015tta}, in this limit, the classical  
scattering cross-section of a graviton moving in the brane background vanishes indicating that the bulk gravity possibly
gets decoupled from the brane. This phenomenon is quite similar to the BPS case \cite{Alishahiha:2000qf}. Now in order to 
find the decoupled geometry we make the following change of variables in analogy with BPS D3 brane 
\cite{Maldacena:1997re,Aharony:1999ti}, 
\be\label{decoupling}
\rho=\alpha'u, \quad \rho_0=\alpha'u_0,\quad\a \cosh^2\theta = \frac{L^4}{u_0^4\alpha'^2}
\ee
along with $\a' \to 0$. Note that in the above $u$ and $u_0$ have the dimensions of energy and are kept fixed
as we take $\alpha' \to 0$. Also $L^4 = 2 g_{\rm YM}^2 N = \frac{R^4}{\a'^2}$ is a dimensionless parameter and remains fixed, 
where, $g_{\rm YM}^2 N$ is the 't Hooft coupling of the boundary theory.  
We would like to point out that in the limit, as $\alpha'\to 0$, $\rho_0\to 0$ and $\theta\to \infty$, but, that does not 
imply that we have the BPS 
limit. This is because here $\rho$ and $\rho_0$ go to zero with the same scale and therefore, $G(\rho)$ does not go to 1 
as in the BPS limit. Furthermore, note from \eqref{decoupling} and \eqref{constraints} that 
in the limit $\a' \to 0$, the charge has the value $Q/\a'^2 \sim L^4 \gg 1$. The last relation follows
from the fact that the curvature of space-time in string units must be very very small for the supergravity
description to remain valid. 
In \cite{Nayek:2015tta} we found that the decoupling must occur also for small or even zero charge 
of the non-susy D3 brane, but we have not been able to find the explicit decoupling limit for these cases.  

Now to justify the decoupling limit \eqref{decoupling}, we will see how in this limit we can keep the energy of a particle 
in the throat region in string units as well as that measured by an observer at infinity fixed \cite{Aharony:1999ti}. Since 
$g_{tt}$ as given in
\eqref{nonsusyd3n} is not constant these two energies will not be the same.
So, if $E_p$ denotes the energy of a particle as measured by an observer at a finite distance $\rho$ from the brane 
and $E$ denotes that of the same particle as measured by an observer at infinity, then they are related by a red-shift
factor given by,
\be\label{energy}
E=\sqrt{g_{tt}}E_p=F(\rho)^{-\frac{1}{4}}G(\rho)^{\frac{\delta}{8}}E_p
\ee  
Under the decoupling limit \eqref{decoupling} the functions $G(\rho)$ and $F(\rho)$ become
\bea\label{hfunctions}
&& G(\rho)\rightarrow G(u)=1+\frac{u_0^4}{u^4} = {\rm fixed}\nn
&& F(\rho)\rightarrow F(u)=\tilde F(u)\frac{L^4}{\a u_0^4\alpha'^2}
\eea
where, $\tilde F(u)=G^{\frac{\alpha}{2}}(u)-G^{-\frac{\alpha}{2}}(u)$. Then \eqref{energy} takes the form
\be\label{energy1}
E=\tilde F(u)^{-\frac{1}{4}}\frac{\a^{\frac{1}{4}}u_0}{L}G(u)^{\frac{\delta}{8}}(\sqrt{\alpha'}E_p)
\ee
Therefore, if we keep $\sqrt{\a'}E_p$ fixed, then $E$ will remain fixed since the other quantities on the rhs of \eqref{energy1} are
fixed in the decoupling limit. This gives a consistency check of the decoupling limit with the energy of an arbitrary excited string
state and in turn implies (alongwith our observation in \cite{Nayek:2015tta}) that the low energy excitations near $\rho=0$ and 
those in the bulk get decoupled in the decoupling limit. 
We can recover the results for the BPS D3 brane from here by putting $u_0 \to 0$. Now, $G(u) \to 1$ and 
$F(u) = {\tilde F}(u)\frac{L^4}{\a u_0^4 \a'^2} \to \frac{L^4}{\a'^2 u^4}$ and the energy relation \eqref{energy1} reduces to 
$E = \frac{u}{L}(\sqrt{\a'}E_p)$, precisely that of a BPS D3 brane \cite{Aharony:1999ti}. 

{\em Throat geometry} -- Here we will discuss the spacetime geometry for the 
non-susy D3 brane in the decoupling limit 
\eqref{decoupling}  we discussed in the previous section. In case of 
BPS D$3$ brane, the background becomes AdS$_5\times$S$^5$ in the corresponding decoupling limit.
We have seen the form of the various functions in the decoupling limit in \eqref{hfunctions}. The non-susy D3 brane solution 
\eqref{nonsusyd3n} in the string frame becomes
\bea\label{throat}
ds^2 & = & \alpha'\frac{L^2}{u_0^2}\Bigg[\tilde F(u)^{-\frac{1}{2}}G(u)^{\frac{\delta}{4}}
\left(-dt^2+\sum_{i=1}^3(dx^i)^2\right)\nn
&&+\tilde F(u)^{\half}G(u)^{\frac{1+\delta}{4}}\left(\frac{du^2}{G(u)}+u^2d\Omega_5^2\right)\Bigg]\nn
e^{2\phi} & = & g_s^2G(u)^\delta, \quad F_{[5]} = \frac{2\sqrt{2} \a'^2}{\kappa}L^4 (1+\ast){\rm Vol}(\Omega_5) 
\eea
Here we have restored the string coupling constant $g_s$, and $\kappa = \sqrt{8\pi G_{10}}$, where
$G_{10}$ is the ten dimensional Newton's constant. The Yang-Mills coupling constant is related to 
$g_s$ by $g_{YM}^2=2\pi g_s$ and is 
independent of $\alpha'$. Also in the above we have redefined the coordinates 
$(t,\, x^i) \to \frac{L^2}{\sqrt{\a} u_0^2} (t,\,x^i)$, for $i=1,\,2,\,3$ and rescaled $L^2 \to \sqrt{\a}L^2$.
The effective string coupling constant $e^{\phi} = \frac{g_{\rm eff}^2}{N} = g_s G(u)^{\frac{\d}{2}} = \frac{g_{\rm YM}^2}{2\pi} G(u)^{\frac{\d}{2}}$ 
is also independent of $\alpha'$. We, therefore, claim \eqref{throat} to be the throat geometry of non-susy D3 brane. It can be easily
checked that in the BPS limit $u_0 \to 0$ the above geometry reduces to AdS$_5$ $\times$ S$^5$. The same geometry can also be
obtained in the asymptotic limit, i.e., for $u \to \infty$. Now since there is decoupling of gravity on the non-susy D3 brane, this
geometry must be dual to a QCD-like theory. To see that this is indeed true we will map the decoupled geometry
\eqref{throat} to the previously known geometry given by Constable and Myers \cite{Constable:1999ch} quite a while ago. In order 
to do that we redefine the function $\tilde F(u)$ as
$\tilde F(u)={\hat F}(u) G(u)^{-\frac{\alpha}{2}}, ~~ {\rm where,} ~ {\hat F}(u)=G(u)^\alpha -1$.
The metric in the Einstein frame and the dilaton then take the forms,
\bea\label{metricdilaton}
ds^2 &=& \alpha'\frac{L^2}{u_0^2}\Bigg[{\hat F}(u)^{-\frac{1}{2}}G(u)^{\frac{\alpha}{4}}\left(-dt^2+\sum_{i=1}^3 (dx^i)^2\right)\nn
&& +{\hat F}(u)^{\half}G(u)^{\frac{1-\alpha}{4}}\left(\frac{du^2}{G(u)}+u^2 d\Omega_5^2\right)\Bigg]\nn
e^{2\phi} & = & g_s^2G(u)^\delta
\eea
Then we make a coordinate transformation
\be\label{coordinatechange} 
u=\bar{r}\left(1+\frac{u_0^4}{4\bar{r}^4}\right)^{-\frac{1}{4}} \equiv \bar{r}\left(1+\frac{\omega^4}{\bar{r}^4}\right)^{-\frac{1}{4}}.
\ee
So the old harmonic function is modified to $G(u) \to (1+2\omega^4/\bar{r}^4)^2$ and $\hat{F}(u) \to (1+2\omega^4/\bar{r}^4)^{2\a} -1
\equiv \hat{H}(\bar{r})$. Therefore we find that, the metric and the dilaton \eqref{metricdilaton} in this new coordinate, 
matches exactly with the Constable-Myers solution eqn.(2.1) of \cite{Constable:1999ch} if we identify the
parameters as $\a = \d_{\rm CM}/2$ and $\d = \Delta_{\rm CM}/2$, where we have denoted the Constable-Myers parameters with a subscript `CM'.
The parameter relation $\a^2+\d^2=5/2$ given in \eqref{constraints} then becomes $\d_{\rm CM}^2 + \Delta_{\rm CM}^2 = 10$ and is precisely
the parameter relation given in Constable-Myers solution. We thus claim that the throat geometry in the decoupling limit of the non-susy 
D3 brane solution is nothing but the Constable-Myers two parameter solution. Since we already found that the bulk gravity gets decoupled
for the non-susy D3-brane in the decoupling limit, so, the throat geometry must be the gravity dual of some QCD-like theory as discussed
by Constable and Myers and our calculation justifies that.

We remark that the Constable-Myers two parameter solution have also been shown in \cite{Constable:1999ch} to arise from a 
suitable scaling limit of
a non-susy D3 brane solution very similar to the decoupling limit we have discussed. But how that scaling ($\beta$) is related to the physical
low energy limit ($\a' \to 0$) is not clear there. Identifying $\beta = \cosh\theta$ in our solution we find from \eqref{decoupling} that 
$\beta = L^2/(\sqrt{\a} u_0^2 \a')$ which indeed goes to infinity in the decoupling limit $\a' \to 0$. This clarifies why their
scaled solution decouples gravity and represents a gravity dual of Yang-Mills type theory, actually a deformation of $D=4$,
${\cal N}=4$ super Yang-Mills theory, by breaking susy and conformal symmetry. This theory also contains massive fermions and 
scalars in the adjoint representation and has been shown to exhibit various QCD-like properties such as running coupling, confinement 
and mass gap in the glueball spectrum in certain range of parameters. Asymptotic bahavior of the dilaton and the volume scalar determine 
the expectation values of the gauge invariant dimension four Tr$(F^2)$ and dimension eight Tr$(F^4 - (F^2)^2))$ operators in the 
gauge theory \cite{Constable:1999ch} in terms of the parameters of the gravity theory.                    

We will now show that the decoupled gravity background of non-susy D3 brane \eqref{metricdilaton} describing QCD-like theory also studied 
by Constable and Myers can be reduced, in a special case, to another gravity background, supposed to describe infrared QCD-like theory which 
includes non-perturbative gluon condensate providing a natural IR cut-off for confinement, studied by Csaki and Reece \cite{Csaki:2006ji}. For 
this we put $\a=1$.   
So, by the constraint relation \eqref{constraints} we have $\d = \pm \sqrt{\frac{3}{2}}$. We will take only the negative sign 
for $\d$ because as $u \to 0 $,
we want to keep the dilaton small. Also, $\hat{F}(u)$ now takes the form  
$\hat{F}(u)=G(u)-1=u_0^4/u^4$.
The metric and the dilaton \eqref{metricdilaton} then reduce in a new coordinate $z=\frac{L^2}{u}$ to,
\bea\label{newmetricdilaton}
&& ds^2=\alpha'\left[\frac{L^2}{z^2}\left(G(z)^{\frac{1}{4}}\big(-dt^2+\sum_{i=1}^3(dx^i)^2\big) + \frac{dz^2}{G(z)}\right)\right.\nn
&& \qquad\qquad\qquad\qquad \left. +L^2d\Omega_5^2\right]\nn
&& e^{2\phi}=g_s^2G(z)^{-\sqrt{\frac{3}{2}}}
\eea
where $G(z) = 1 + \frac{z^4 u_0^4}{L^8} \equiv 1+ \frac{z^4}{z_0^4}$.
Note that the metric in \eqref{newmetricdilaton} asymptotically
($z \to 0$) has the form AdS$_5$ $\times$ S$^5$. To cast the metric and the dilaton into the form of Csaki and Reece we need to go to
another coordinate given by,
\be\label{anothercoordinate}
\hat{z}=z\left(\frac{1+\sqrt{G(z)}}{2}\right)^{-\half}  
\ee
where $G(z)$ is as given above. Now using \eqref{anothercoordinate} one can check the following identities,
\bea\label{harmonic1}
& & \frac{2\sqrt{G(z)}}{1 + \sqrt{G(z)}} = 1 + \frac{\hat{z}^4}{\hat{z}_0^4} \equiv  H(\hat{z})\nn
& & \frac{2}{1 + \sqrt{G(z)}} = 1 - \frac{\hat{z}^4}{\hat{z}_0^4} \equiv \tilde{H}(\hat{z})
\eea
where in the above we have defined $\hat{z}_0 = \sqrt{2} z_0$. Now using \eqref{harmonic1}, we can express \eqref{newmetricdilaton} in
terms of the harmonic functions $H(\hat{z})$ and $\tilde{H}(\hat{z})$ and find that they match exactly with the 
metric and the dilaton obtained by Csaki and Reece (eq.(3.10) and eq.(3.11) of \cite{Csaki:2006ji}) as a gravity dual of a QCD-like theory.
We have shown that this background
is nothing but a special case of the throat limit of non-susy D3 brane. As we have seen in \cite{Nayek:2015tta} that the bulk 
gravity indeed gets 
decoupled from the non-susy D3 brane in the decoupling limit, so our calculation justifies the gauge theory (or QCD-like theory)
interpretation of this gravity background.   

To conclude, in this Letter we have obtained the decoupling limit and the throat geometry of the
well-known non-susy D3 brane solution of type IIB string theory.     
We would like to point out that the decoupling limit we have obtained in this Letter is only for the non-susy D3 brane with
large RR charge. However, as we have seen in our previous work \cite{Nayek:2015tta} that decoupling occurs also for the zero charge non-susy
D3 brane. But we have not been able to obtain the decoupling limit for the zero charge case. Since BPS branes are always 
charged, we can take a guidance from it to obtain the decoupling limit for the charged non-susy D3 branes, as we have done
in this paper, but there is no such analog for the chargeless case. However, it will certainly be interesting to understand 
the decoupling limit for the chargeless case and it remains an open problem.
It would be interesting to use gauge/gravity duality to further explore the properties of the QCD-like theory that can be
obtained from the throat geometry of the non-susy D3 brane background. For example, it may be possible to calculate the Wilson
loop, static as well as velocity-dependent quark-antiquark potential, screening length, monopole-antimonopole potential, jet quenching
parameter and compare with the results already known for the susy ${\cal N}=4$ gauge theory.    

\vs{20}

{\em Acknowledgement} -- One of us (KN) wants to thank Arnab Kundu for helpful discussions.

\end{document}